\documentstyle[prd,aps,preprint,psfig,floats]{revtex}

\newcommand{\Mtau}{{M}_{\tau}}
\newcommand{\be}{\begin{equation}}
\newcommand{\ee}{\end{equation}}
\newcommand{\beqn}{\begin{eqnarray}}
\newcommand{\eeqn}{\end{eqnarray}}

\tightenlines
\draft

\begin{document}

\preprint{OKHEP--97--01}

\title{
 Analytic perturbation theory and inclusive $\tau$ decay  }

\author{
K. A. Milton,$^a$\thanks{E-mail: milton@mail.nhn.ou.edu}
I. L. Solovtsov,$^{b}$\thanks{E-mail: solovtso@thsun1.jinr.dubna.su}
and O. P. Solovtsova$^{b}$\thanks{E-mail: olsol@thsun1.jinr.dubna.su} }

\address{
         $^a$Department of Physics and Astronomy,
         University of Oklahoma, Norman, OK 73019 USA\\
         $^b$Bogoliubov Laboratory of Theoretical Physics,
         Joint Institute for Nuclear Research,\\
         Dubna, Moscow Region, 141980, Russia
	}
\date{\today}

\maketitle

\begin{abstract}
We apply analytic perturbation theory to the
inclusive decay of a $\tau$~lepton into hadrons.
It is shown that the resulting analyticity of the coupling constant
strongly influences the value of the QCD $\Lambda$-parameter extracted
from the experimental data on $\tau$ decay.

\end{abstract}


\section{Introduction}

The inclusive character of the decay of a $\tau$ lepton into hadrons and
the fact that the nonperturbative QCD contributions to this process
are very small~\cite{[1]} make it possible, in principle,  to describe this
process on the basis of the standard methods of quantum field theory
without any model assumptions.
Measurement of the ratio of hadronic to leptonic
$\tau$ decay widths, i.e., the quantity
$R_{\tau}\, =\, {\Gamma (\,\tau \to
{\rm hadrons}} +\nu\,)/{\Gamma (\,\tau \to l\nu\overline{\nu}\,)}
\>$,
allows one to extract, with a high degree of accuracy, the value
of the strong coupling constant $\alpha_{\rm QCD}$ at the $\tau$ mass,
$Q=M_\tau\simeq1.78$~GeV.
Comparing this value with the values of $\alpha_{\rm QCD}$ found
at higher energy, for example, $\alpha_{\rm QCD}(Q^2={M}^2_Z)$, is
an important test of the applicability of QCD
perturbation theory over a wide range of energies
and requires a careful check.
At present, both experimental and theoretical investigations of
the decay of a ${\tau}$ lepton are continuing intensively
(see the recent reviews in~\cite{[2]}).

One usually employs analytic properties of the hadronic correlation function
in order to rewrite the original expression for the $\tau$ hadronic rate,
which involves integration over a nonperturbative region of small momenta,
in the form of a contour integral over a circle of sufficiently large
radius $Q^2=M_\tau^2$ to apply perturbation theory (PT)~\cite{[1],[2]}.
However, the perturbative approximation, which introduces a ghost pole,
violates the analytic properties required to use Cauchy's theorem in this
manner.

In this paper, we will apply analytic perturbation theory (APT)~\cite{ss1,ss2}
in which it is possible to maintain the correct analytic behavior.
Within this approach the correct analytic properties of the running coupling
are provided by nonperturbative contributions which emerge automatically
from dispersion relations.\footnote{Some other aspects of the
dispersive approach have been discussed in~\cite{dmw,dw,dkt} and in the
recent papers~\cite{grunberg}. }
The analytic running coupling constant obtained
in such a way turns out to be remarkably stable for the whole interval of
momentum, and has a universal infrared limit at $Q^2=0$, independent of
the value of the QCD scale parameter $\Lambda$.

For our purpose, it is important that, within this approach, it is possible
to give a self-consistent definition of the running coupling constant in the
Minkowskian (timelike) region~\cite{MS1}.  This fact allows us to
obtain two equivalent representations for the QCD correction to $\tau$ decay,
involving the timelike and the spacelike definitions of the running coupling
constants, respectively.  For our numerical estimations we will use APT at
2-loop order.

\section{Why is analyticity important? }

The initial theoretical expression for $R_\tau$ in the case of massless quarks
contains an integral over timelike momentum~$s$~~\cite{[1]}
\be\label{Eq.1}
R_{\tau} =  \frac{2}{\pi}\;  \int^{M_\tau^2}_0 {ds \over M_\tau^2 } \,
\left(1-{s \over M_\tau^2}\right)^2
\left(1 + 2 {s \over M_\tau^2}\right)
{\rm Im}\, \Pi(s) \;,
\ee
where the range of integration extends down to small $s$ and cannot be
calculated in the framework of the standard
perturbation theory. The method of calculation of $R_{\tau}$
based on exploiting certain analytic properties of the hadronic
correlators $\Pi(s)$ allows one to rewrite the expression~(\ref{Eq.1})
by using the Cauchy theorem in the form of a contour integral in the
complex $s$-plane with the contour running counterclockwise around a circle
centered on the origin of radius $M_{\tau}^2$:
\begin{equation}
\label{Eq.2}
R_{\tau}\, =\,
\frac{1}{2\pi {\rm i}} \,
\oint_{|s|=M_\tau^2}
\frac{ds}{s}\,
\left(1-{s \over \Mtau^2}\right)^3 \left(1 + {s \over \Mtau^2}\right)
D(s) \, ,
\ee
where\footnote{Here, we will use the standard definition $q^2=-Q^2$,
so that in the Euclidean region $Q^2>0$.}
$$ D(q^2)\,  =\,  - q^2 \,\frac{d\,\Pi(q^2)}{d\,q^2}\; .$$

In the representation (\ref{Eq.2}) the contour has a sufficiently large
radius, and it is possible, in principle, to calculate $R_{\tau}$
perturbatively. However, the transition to the contour
representation requires certain analytic properties of the correlator.
Namely, the correlator $\Pi(s)$ is an analytic function in the
complex $s$-plane with a cut along the positive part of the real axis.
The parametrization of $\Pi$ by the perturbative running
coupling constant violates these analytic properties~\cite{[3]}.
It is clear that the difference in the regions of integration in
the initial expression~(\ref{Eq.1}) for $R_{\tau}$ and the
expression~(\ref{Eq.2}) obtained after applying the Cauchy theorem
makes it  necessary to parametrize
$\Pi$ in Eq.~(\ref{Eq.1}) and $D$ in Eq.~(\ref{Eq.2}) with different
coupling constants. Indeed, a renormalization-group analysis gives a running
coupling constant determined in the spacelike (Euclidean) region, while the
initial expression (\ref{Eq.1}) contains an integration over timelike
momentum and therefore to calculate Eq.~(\ref{Eq.2}) requires some procedure
of analytic continuation from spacelike to timelike momentum.
To this end, we will determine the effective coupling constant in the
spacelike region ($t$-channel), $\bar{a}^{\rm eff}$, and in the timelike
region, ${\bar{a}}_s^{\rm eff}$, (using the definition that
$a=\alpha_{\rm QCD}/4\pi$)  as (where the overbar signifies that the coupling
constant has the correct analytic properties)
\begin{eqnarray}
\label{Eq.3}
& & D(q^2) \sim  1\,+\,d_1\,a(q^2)\,+\,d_2\,a^2(q^2)\,+ \cdots \,
=\,1\,+\,d_1\,\bar{a}^{\rm eff}(q^2)\, , \\
\label{Eq.4}
& &{\rm Im} \,\Pi(s)
\sim  1\,+\,r_1\,a_s(s)\,+\,r_2\,a_s^2(s)\,+\cdots
\,=\,1\,+\,r_1\,{\bar{a}}_s^{\rm eff}(s)\, ,
\end{eqnarray}
where the first perturbative coefficients $d_1$ and $r_1$ are equal
to each other: $d_1=r_1=4$.

The dispersion relations for the $D$-function yield the following
relation between the effective coupling constants:
\begin{eqnarray}
\label{Eq.5}
\bar{a}^{\rm eff}(q^2)&=&-\,q^2\,\int_0^{\infty}\,
\frac{ds}{{(s-q^2)}^2}\,\, \bar{a}_s^{\rm eff}(s)\, ,\\
\label{Eq.6}
\bar{a}_s^{\rm eff}(s)&=&-\,\frac{1}{2\pi {\rm i}}\,
\int _{s-{\rm i}\,\epsilon} ^{s+{\rm i}\,\epsilon} \frac{dz}{z}\,\,
\bar{a}^{\rm eff}(z)\, .
\end{eqnarray}
Now, separating the QCD contribution $\Delta_\tau$ in $R_\tau$,
\begin{equation}
\label{Eq.7}
R_\tau = 3 \; (|V_{ud}|^2 + |V_{us}|^2)\;S_{{\rm EW}}\;
 ( 1 + \Delta_\tau ) \; ,
\end{equation}
where $V_{ud}$ and $V_{us}$ are the CKM matrix elements, and
$S_{{\rm EW}}$ is the electroweak factor (see~[1]),
we obtain from
Eqs. (1) and (2) the two equivalent representations,
\begin{equation}
\label{Eq.8}
\Delta_\tau \,=\, 2\,r_1\, \int^{M_\tau^2}_0 {ds \over M_\tau^2 } \,
\left(1-{s \over M_\tau^2}\right)^2
\left(1 + 2 {s \over M_\tau^2}\right) \bar{a}_s^{\rm eff}(s)\;,
\end{equation}
and
\begin{equation}
\label{Eq.9}
\Delta_\tau \,=\,\frac{d_1}{2\pi {\rm i}} \, \oint_{|z|=M_\tau^2}
\frac{dz}{z}\,
\left(1-{z \over \Mtau^2}\right)^3 \left(1 + {z \over \Mtau^2}\right)
\bar{a}^{\rm eff}(z) \,.
\end{equation}

It should be noted that the equivalence of these formulae
holds only in the case of the above-mentioned analytic properties
of the correlator $\Pi(s)$ and the $D$-function and that
these analytic properties are broken in the standard perturbation theory.
The QCD contribution represented in the form~(\ref{Eq.2}) is the expression
which one usually uses for theoretical analysis. In principle,
the expression (\ref{Eq.9}) can be calculated on the basis of perturbation
theory, but then it is impossible to return from Eq.~(\ref{Eq.9}) to
Eq.~(\ref{Eq.8}) which corresponds to the initial Eq.~(\ref{Eq.1}), and
nothing can be said about the error associated with
switching from Eq.~(\ref{Eq.8}) to Eq.~(\ref{Eq.9}). Therefore, it is 
impossible to give in the framework of the standard perturbation theory a 
self-consistent description of the inclusive decay of a $\tau$~lepton into 
hadrons.

\section{The APT method for ${\tau}$-decay}

Now, we apply analytic perturbation theory~(APT), which was recently
proposed in~\cite{ss1,ss2,MS1}, to describe the inclusive decay of
a $\tau$~lepton into hadrons and to give a quantitative estimate of the
effect
associated with $Q^2$-analyticity of the running coupling constant.
The method of APT allows one to avoid the above-mentioned
difficulties and evaluate both the
initial integral over the physical region and the contour representation
which are equal due to the Cauchy theorem.
In the framework of APT, $\Pi$ and the $D$-function can be parametrized by the
running coupling constant with the correct analytic properties
as in Eqs.~(\ref{Eq.3}) and (\ref{Eq.4}), where the running coupling
constants in the space- and timelike regions can be expressed
in the terms of the spectral density $\varrho(\sigma)$ as~\cite{MS1}
\begin{eqnarray}
\label{Eq.10}
\bar{a}(z) &=& \frac{1}{\pi}\,
\int_0^\infty\,\frac{d\sigma}{\sigma\,-\,z\,} \,\varrho(\sigma) \, , \\
\label{Eq.11}
\bar{a}_s(s) &=&\frac{1}{\pi }\,
\int_s^\infty\,\frac{d\sigma}{\sigma}\,\varrho(\sigma)\, .
\end{eqnarray}

In the leading order, the expression for ${\bar{a}}(z)$ has the form
\begin{equation}
\label{Eq.12}
\bar{a}^{(1)}(z)\,=\,
\, \frac{1}{\beta_0}\,
\left[\, \frac{1}{\ln \left(-z/\Lambda^2\right)}\,+\,\frac{1}{1+z/\Lambda^2}
\,\right]\, ,
\end{equation}
where $\beta_0=11-2n_f/3$ is the first coefficient of the $\beta$ function.
The first term in (\ref{Eq.12}) determines the asymptotic behavior at
large momenta and has the same form as in perturbation theory.
The second term, which appears automatically, reproduces the
correct analytic properties, and the ghost pole at
$z=-\Lambda^2$ does not arise. We underscore once again that
APT makes it possible to implement the correct
transition from expression (\ref{Eq.1}) to expression (\ref{Eq.2}).
These expressions simply coincide, as they should, while the application
of perturbation theory with the standard renormalization-group refinement
runs into serious difficulties.

The fundamental quantity in the APT approach is the spectral density
$\varrho(\sigma)$ by which one can parametrize both the running coupling
constants in the spacelike and in the timelike regions. We now
find a formula which expresses the strong interaction contribution
to $R_{\tau}$ via the spectral density function.
To this end, introduce an effective spectral density $\rho^{\rm eff}(\sigma)$
that corresponds to the effective coupling constant [see Eq.~(\ref{Eq.3})]
and write down the effective coupling constants in the Euclidean and physical
regions as follows,
\begin{eqnarray}
\label{a-rho-t}
\bar{a}^{\rm eff}(z) &=& \frac{1}{\pi}\,
\int_0^\infty\,\frac{d\sigma}{\sigma\,-\,z\,} \,\rho^{\rm eff}(\sigma)\, ,\\
\label{a-rho-s}
\bar{a}_s^{\rm eff}(s) &=&\frac{1}{\pi }\,
\int_s^\infty\,\frac{d\sigma}{\sigma}\,\rho^{\rm eff}(\sigma)\, .
\end{eqnarray}

First, consider the contour representation (\ref{Eq.9}) for the QCD
contribution $\Delta_{\tau}$. By using Eq.~(\ref{a-rho-t}) this formula
can be rewritten as follows
\begin{equation}
\label{d-rho}
\Delta_\tau\, =\, \frac{d_1}{\pi}
\int_0^\infty\,\frac{d\sigma}{\sigma}\,\rho^{\rm eff}(\sigma)\, -\,
\frac{d_1}{\pi}
\int_0^{M^2_{\tau}}\,\frac{d\sigma}{\sigma}\,
{\left(1-\frac{\sigma}{M^2_{\tau}}\right)}^3\,
{\left(1+\frac{\sigma}{M^2_{\tau}}\right)}\,\rho^{\rm eff}(\sigma)\, .
\end{equation}
Of course, the same result~(\ref{d-rho}) will be obtained if we substitute
Eq.~(\ref{a-rho-s}) into Eq.~(\ref{Eq.8}).

Due to the feature of universality, the first term in Eq.~(\ref{d-rho})
can be expressed in terms of only the first $\beta$-function
coefficient.
To demonstrate this here let us use the following asymptotic representation
for the effective coupling constant
\begin{equation}
\label{asy-eff}
a_{\rm PT}^{\rm eff}(q^2)\,=\,\frac{1}{\ell}\,\sum_{k=0}^\infty
\sum_{m=0}^k \alpha_{k,m}\frac{\ln^m\ell}{\ell^k}\, ,
\end{equation}
where $\ell=\ln(Q^2/\Lambda^2)$ and $\alpha_{0,0}=1/\beta_0$.
Calculating the effective spectral density as the discontinuity of the
effective coupling (\ref{asy-eff}), for the
infrared limiting value of the analytic effective coupling constant
we get, with $L=\ln\sigma/\Lambda^2$,
\begin{eqnarray}
\label{anal-infra}
\bar{a}_{\rm APT}^{\rm eff}(0)\,&=&\,\frac{1}{\pi}\int_{-\infty}^\infty
d L\, \rho^{\rm eff}(L) \nonumber \\
&=&\,\frac{1}{\beta_0}\,+\,\sum_{k=1}^{\infty}\,\sum_{m=0}^k\,
\alpha_{k,m}\,\Delta\,\bar{a}_{k,m}^{\rm eff}(0)
\end{eqnarray}
with
\begin{equation}
\label{del-aeff}
\Delta\,\bar{a}_{k,m}^{\rm eff}(0)\,=\,\frac{1}{\pi}{\rm Im}
\int_{-\infty}^\infty \,d L\,
\frac{\ln^m (L-{\rm i}\pi)}{(L-{\rm i}\pi)^{k+1}}\, .
\end{equation}
By repeatedly integrating by parts, it is now very easy to show that
\begin{equation}
\label{daeffzero}
\Delta \,\bar{a}_{k,m}^{\rm eff}(0)\,=\,\frac{m!}{k^m}\Delta\,
\bar{a}_{k,0}^{\rm eff}(0)\,=\,0.
\end{equation}

Consider the 2-loop level, in which the perturbative effective running
coupling constant is
\begin{equation}
\label{pt-eff}
a^{\rm eff}_{\rm PT}(q^2)\,=\,\bar{a}_{\rm PT}(q^2)\,+\,
\frac{d_2}{d_1}\bar{a}_{\rm PT}^2(q^2) \, ,
\end{equation}
where for $n_f$ active quarks $d_2=16\,(1.9857-0.1153n_f)~\cite{gorishny}$.
By using Eq.~(\ref{pt-eff}) we can obtain the corresponding effective
spectral density.
It is important to appreciate that in the framework of the APT approach
there is very little sensitivity to the approximation used in solving
the renormalization group equations.  This fact is demonstrated
in Fig.~\ref{rho}, where we plot three spectral densities.
The solid line corresponds to the spectral density obtained from the exact
quadrature of the 2-loop $\beta$ function,
\begin{equation}
\label{exactrg}
\beta_0\ell\,=\,\int^{\bar{a}}\frac{dx}{\beta(x)}=\frac{1}{\bar{a}}
-\beta_0B_1\ln\left(1+\frac{1}{\beta_0B_1\bar{a}}\right) \, ,
\end{equation}
where $B_1=\beta_1/(\beta_0)^2$, and
$\beta_1=102-38n_f/3$ is the two-loop coefficient of the
$\beta$-function.

The dashed line corresponds to the spectral density obtained from the
asymptotic representation of the perturbative 2-loop running coupling,
\begin{equation}
\label{asymrg}
\bar{a}\,=\,\frac{1}{\beta_0\ell}\left(1-B_1\frac{\ln\ell}{\ell}\right)\,.
\end{equation}

	        \begin{figure}[hbt]
        
        \centerline{
\psfig{figure=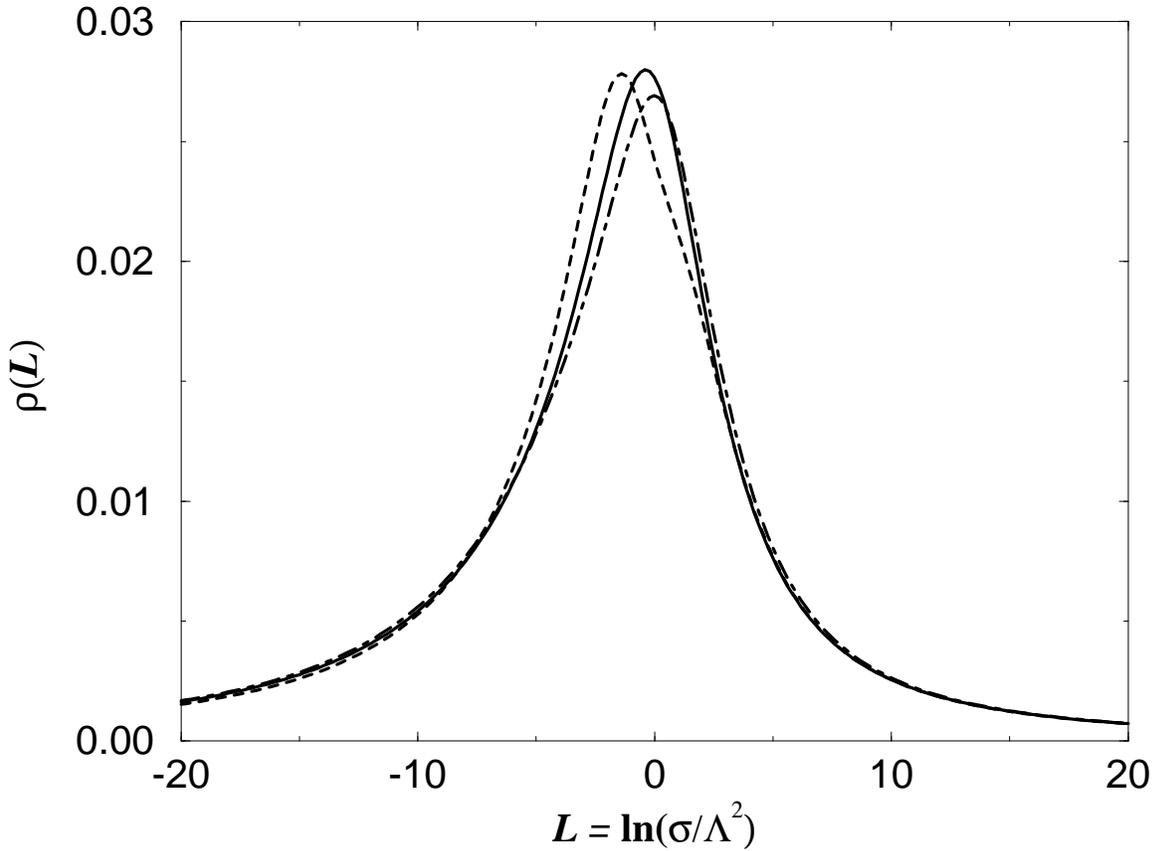,height=5in,width=6.5in,angle=270}}
\vskip1cm
	  \caption{  {\sl Plot of the spectral densities.
	  The solid line refers to the exact quadrature of the RG
	  equation, the dashed line refers to the asymptotic representation
	  from the perturbative running coupling, and the dot-dashed line
	  corresponds to solution obtained by one iteration.
	  			        }     }
	         \label{rho}
          \end{figure}

The dot-dashed line corresponds to the following perturbative running
coupling
\begin{equation}
\label{af2}
\bar{a}(Q^2)=\frac{1}{\beta_0 \ell + \beta_0 B_1 \ln (1+\ell/B_1)}~ ,
\end{equation}
This expression, which we will use for our numerical calculations,
is the result of exact integration of the two-loop differential RG
equation (\ref{exactrg}) solved by one iteration.
For this case the spectral density which is associated with the running
coupling (\ref{af2}) is
\begin{equation}
\label{rho2}
\varrho(\sigma)\,=\,\frac{1}{\beta_0 }\,
\frac{I(L)}{R^2(L)\,+\,I^2(L)}\, ,
\end{equation}
with
\begin{eqnarray}
\label{ri}
R(L)\,&=&\,L\,+\, B_1\ln \sqrt{\left(1+\frac{L}{B_1}\right)^2+
\left(\frac{\pi}{B_1} \right)^2}~, \\ \nonumber
I(L)\,&=&\,\pi\,+\, B_1{\rm arccos}\frac{B_1+L}
{\sqrt{\left(B_1+L\right)^2+\pi^2}}~.
\end{eqnarray}
As a result, for the effective spectral density, we find
\begin{equation}
\label{eff-rho}
\rho^{\rm eff}(\sigma)\,=\,\varrho(\sigma)\,+\,
\frac{1}{\beta_0^2}\frac{d_2}{d_1}\frac{2R(L)I(L)}{[R^2(L)+I^2(L)]^2}\, .
\end{equation}

By substituting this formula into Eq.~(\ref{d-rho}) and using the fact of
universality for the first term we obtain the QCD correction for the
inclusive decay rate of the $\tau$ lepton. For our numerical estimations
we will use the world average $R_{\tau}=3.633\pm0.031$~\cite{PDG96}, which
leads to the following values of the APT running couplings:
$\alpha_s(M_{\tau}^2)=0.378\pm 0.026$ in the timelike region and
$\alpha(M_{\tau}^2)=0.400\pm 0.026$ in the spacelike region, respectively.
The reader should carefully note that the values of the strong coupling
constant in the timelike and the spacelike region do not agree, even though
they are evaluated at the same scale.  This difference could become
significant experimentally as the precision of the measurements improves.

It should be stressed that these values of the coupling constants have been
obtained with a value of the QCD scale parameter much
larger\footnote{The same observation has been made in the one-loop
level in~\cite{olga}.}
than that
found in conventional perturbation theory: for three active flavors,
$\Lambda_{(3)}=935\pm159$ MeV.
We have found that
the value of $\Lambda$ is very sensitive to the value of $R_\tau$.
For example, if we were to use a smaller value of
$R_\tau=3.559\pm0.035$~\cite{CLEO}, we would obtain
$\Lambda=640\pm 127$ MeV.
Note here that, as we have demonstrated, the conventional perturbative
parameterization of the $D$-function is inconsistent with the required
analyticity properties and it is not possible to write the contour integral
(\ref{Eq.2}) in order to extract $\Lambda_{\rm QCD}$.
Nevertheless, in order to give the reader a feeling for the relation between
the APT and PT values of $\Lambda$, we require that that the 2-loop
$D$-functions be the same in both schemes at the $\tau$ scale, and obtain
the relation between the two definitions of $\Lambda$ shown in
Fig.~\ref{l-pt-apt}.

	        \begin{figure}[hbt]
         \centerline{
\psfig{figure=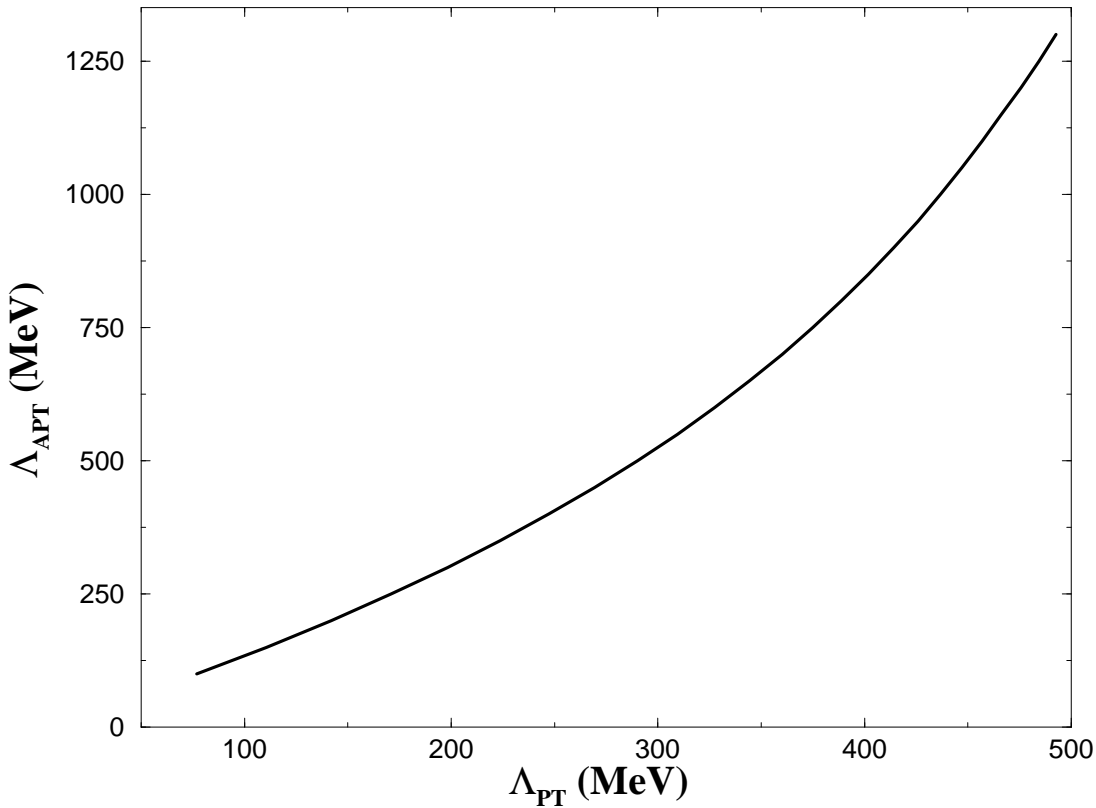,height=5in,width=6.5in,angle=270}}

	  \caption{  {\sl $\Lambda_{\rm APT}$  vs. $\Lambda_{\rm PT}$
	  at the $\tau$-mass scale.
			        }     }
	         \label{l-pt-apt}
          \end{figure}

It is possible to understand in a simple manner why in this approach the
$\Lambda$ parameter is so much larger than the PT value.  To this end,
we will use the approximate formula for the 2-loop APT coupling constant
\cite{ss2}:
\begin{equation}
\label{aapt2}
\bar{a}(q^2)\,\simeq\,\bar{a}_{\rm PT}(q^2)+\frac{1}{2\beta_0}
\frac{1}{1+q^2/\Lambda^2}+\frac{C_1}{\beta_0}\frac{\Lambda^2}{q^2}\;,
\end{equation}
where, for three active quarks, $C_1=0.035$. The accuracy of this formula
on the interval $2.5<|q|/\Lambda<3.5$ is about 0.5\%, and Eq.~(\ref{aapt2})
practically coincides with the exact formula for larger values of momentum.
A key feature of this formula is that the perturbative and nonperturbative
contributions are separated, as in the 1-loop case, which allows us to write
the QCD correction to $\tau$ decay as the sum of perturbative and
nonperturbative terms.  The main part of the corresponding nonperturbative
contribution is
\begin{equation}
\label{deltanp}
\Delta_{\rm NP}\,=\,-\frac{d_1}{\beta_0}\left(1+2\frac{d_2}{d_1}
\frac{1}{\beta_0}\right)\frac{\Lambda^2}{M_\tau^2}\;.
\end{equation}
Due to the negative sign of this contribution, the perturbative term must be
larger than in the case of PT (which implies a larger value of $\Lambda$),
in order to obtain the same value of the QCD correction.

\section{Conclusion}

We have considered the method of analytic perturbation theory
and its application to the semileptonic decay of the $\tau$ lepton.
It should be stressed that the principle of causality, the consequence of
which are certain analytic properties of the running coupling constant,
leads to the essential stability of the running coupling constant in the
infrared region with respect to higher loop corrections. Here, the prime point
is the universal value of the analytic coupling constant at $Q^2=0$, which
does not depend on the experimental estimates of the QCD scale parameter,
$\Lambda$, nor on the number of loops in which we approximate the running
coupling constant. In other words, families of curves corresponding to
different values of $\Lambda$ and to different numbers of loops have the
common point $\bar{\alpha}(Q^2=0)=4\pi/\beta_0$, and are represented by a
bundle of curves.

Moreover, this approach allows us to give a self-consistent definition
of the running coupling constant in the timelike region.  Thus, in this
approach it is possible to give two equivalent expressions for the QCD
correction to inclusive $\tau$ decay, either in terms of the timelike
coupling, or in terms of the coupling defined for complex Euclidean momentum.
This is not possible in conventional perturbation theory, due to the
breaking of analyticity.

The fact that we are now able to define a consistent timelike coupling
constant enables us to perform matching between regions with different
numbers of active flavors in the physical region, where at least in the
leading order the number of active quarks is an obvious fact associated with
physical thresholds.  As a result of this matching procedure the coupling
constant in the Euclidean region will know about all physical thresholds
through the dispersion relation.

Analytic perturbation theory, in effect, incorporates power corrections
(nonperturbative effects) into perturbation theory in order to secure
the required analytic structure.  We may refer to these as short-distance
perturbative power corrections \cite{grunberg}.  There are indications that the 
ambiguities connected to the asymptotic character of the conventional
perturbative expansion and the ambiguities in the nonperturbative
matrix elements should be similar to each other.  The convergence properties
of the new expansion are different from those of the conventional 
expansion, and empirically the APT seems to be convergent.  Therefore,
the role of the nonperturbative power corrections, which describe
the long-distance dynamics within the standard approach (in the language
of the operator-product expansion, they are associated with quark and
gluon condensates), is undoubtedly changed.  We will consider the 
importance of long-distance nonperturbative power corrections in this
scheme elsewhere.

The results of the analysis performed above demonstrate the
importance of analyticity in the running coupling constant, not only from
the fundamental point of view---a self-consistent theoretical description of
$\tau$ decay---but also from the standpoint of giving a self-consistent
description of the $Q^2$ evolution of the coupling constant and extracting
the parameter $\Lambda_{{\rm QCD}}$ from the experimental data on $\tau$
decay.

\section*{Acknowledgement}
The authors would like to thank D. V.~Shirkov
for interest in this work and for useful comments.
Partial support of the work by the US National Science Foundation,
grant PHY-9600421, and by the US Department of Energy,
grant DE-FE-02-95ER40923, is gratefully acknowledged.

\bibliographystyle{prsty}

\end{document}